\documentclass[aps,prl,reprint,superscriptaddress,showpacs]{revtex4-2}

\usepackage{graphicx}
\usepackage{dcolumn}
\usepackage{bm}
\usepackage{amsmath}
\usepackage{footnote}

\def \link_col{blue}
\newcommand{\gray}{$\gamma$-ray~}

\begin{document}

\title[gc nlines]{Constraining the low-energy cosmic ray flux in the central molecular zone from MeV nuclear deexcitation line observations}
\author{Bing Liu}
\affiliation{Deep Space Exploration Laboratory, Hefei 230088, China}
\affiliation{CAS Key Laboratory for Research in Galaxies and Cosmology, Department of Astronomy, School of Physical Sciences,\\
University of Science and Technology of China, Hefei, Anhui 230026, China}
\affiliation{School of Astronomy and Space Science, University of Science and Technology of China, Hefei, Anhui 230026, China}

\author{Ruizhi Yang}
\email{yangrz@ustc.edu.cn}
\affiliation{Deep Space Exploration Laboratory, Hefei 230088, China}
\affiliation{CAS Key Laboratory for Research in Galaxies and Cosmology, Department of Astronomy, School of Physical Sciences,\\
University of Science and Technology of China, Hefei, Anhui 230026, China}
\affiliation{School of Astronomy and Space Science, University of Science and Technology of China, Hefei, Anhui 230026, China}

\begin{abstract}

Low-energy cosmic rays (LECRs) dominate the ionization in dense regions of molecular clouds in which other ionizers such as UV or X-ray photons are effectively shielded. Thus it was argued that the high ionization rate at the central molecular zone (CMZ) of our Galaxy is mainly caused by LECRs. However, the required LECR flux is orders of magnitude higher than the extrapolation of GeV cosmic ray (CR) flux derived from GeV \gray observations. In this paper, we considered two types of additional LECR components and found that only very soft anomalous CR components can explain such a high ionization rate. This LECR component will inevitably produce MeV nuclear deexcitation lines due to their inelastic scattering with the ambient gas. We calculated the MeV line emission and discussed the detectability of next-generation MeV instruments. We found that future MeV observations can be used to pin down the origin of the high ionization rate in the CMZ. 
   
\end{abstract}
\maketitle

\section{I. Introduction}
\label{sec:intro}
Cosmic rays (CRs) are one of the most important components of the interstellar medium. Those with energy below 1 GeV per nucleon are usually referred to as low-energy CRs (LECRs). LECRs dominate the ionization and heating of dense molecular clouds, where the star forms, and also play an important role in the astrochemistry process \cite[see,e.g.,][]{Padovani2009,Papadopoulos2010,Gabici2022review}.  However, the measurement of LECRs is difficult, especially for LECRs with energy smaller than the production threshold of neutral pions (of about $280~\rm MeV$). LECRs suffer strong ionization loss and propagate slower in the standard propagation model \cite{galprop}. Thus the density of LECRs in principle can be significantly different in different sites in our Galaxy.

In addition to indirect measurements such as the molecular ionization rate and Fe K$\alpha$ line emission \cite[see,e.g.,][]{Indriolo2010,Indriolo2015,Tatischeff2012}, \gray line emission from nuclear deexcitation is regarded as a unique tool for the probe of LECR nuclei, which can provide more abundant information on both accelerated particles and interacting medium \cite[see,e.g.,][]{Ramaty1979, Murphy2009}. Besides preresearch of MeV line emission around individual CR accelerators 
\cite[see,e.g.,][]{Summa2011casA,nlines,nlines_casa},  
there were also several works done for the estimation of LECR-associated MeV line emission from the Galactic center (GC) \cite[e.g.,][]{Dogiel2009,Benhabiles2013,Angelis2018}. However, these estimations covered a much larger region around GC, and more specific structures can be analyzed and discussed regarding an ideal angular resolution ($\lesssim 2^\circ$ at MeV) that may be achieved by next-generation MeV \gray detectors.

In this regard, the central molecular zone (CMZ) in the center of our galaxy is of prime interest to such a study. Extending $\sim$200\,pc from the center (a GC distance of 8 kpc is applied in this paper), the CMZ holds around 10\% of the Milky Way's interstellar molecular mass \cite{Genzel1994}. Dense molecular clouds within the CMZ have particle densities of around $10^4\,{\rm cm}^{-3}$, fostering intense stellar activity and evolution, hosting numerous massive stars and supermassive clusters. It is marked by turbulent interstellar medium with strong magnetic fields, and energetic phenomena like supernova remnants and cataclysmic variables \cite{Morris1996}. The  CR ionization rates measured by astrochemistry methods are significantly higher in the CMZ than in other parts of our Galaxy\cite{Dogiel2015}. This region offers insights into the complex dynamics of gas, magnetic fields, and stellar processes in the GC.

The CMZ has already been observed extensively in the entire electromagnetic radiation band. The high-energy CRs (HECRs, here referred as CRs with kinetic energy $>1$\, GeV 
 per nucleon ) in the CMZ, which produce continuum gamma-ray emission via proton-proton inelastic collisions with the gases have been studied via various gamma-ray observations.
Especially, the TeV \gray observations have revealed that CMZ harbor potential PeV CR accelerator \cite{hess2016gc}, while in the GeV band, the CR density is even lower than in the vicinity \cite{yang14, Huang2021cmz}. Thus a simple extrapolation seems to predict a lower LECR density in this region, and the corresponding ionization rate ($\zeta\sim 10^{-17} {\rm s}^{-1}$) is several magnitudes lower than recent measurements ($\zeta\sim 2\times10^{-14} {\rm s}^{-1}$,\citealt{Oka2019}). Due to the very high gas density in the CMZ and effective shielding of UV and X-ray photons, the most probable source of ionization is the LECRs. The possible tension of the observations in different energy bands can be settled by assuming a new LECR component in the CMZ \cite{Benhabiles2013} or regarding that the high ionization rates are only valid for the surface of the molecular clouds \cite{Dogiel2015}. In this paper, we calculated the MeV nuclear deexcitation line of LECRs in the CMZ. We found that the MeV line measurement with future MeV detectors can directly measure the CR density and distinguish the origin of the high ionization rate in CMZ, which would induce a significant impact in understanding the star-forming processes in the GC.

This paper is organized as follows. 
In Sec.II, under the constraint of GeV-TeV gamma-ray observation, we describe the spectra distribution of CR combined with the assumed low-energy components in the CMZ. Next, in Sec.III, we calculate the corresponding ionization rate and 6.4-keV Fe K$\alpha$ line emission caused by CRs in the CMZ. Then in Sec.IV, we estimate the MeV nuclear deexcitation line emission generated by the interactions among CR nuclei and the gases in the CMZ. Finally, we discuss and summarize the above results in Sec.V.

\section{II. Spectral distribution and compositions of CRs in CMZ}
\label{sec:cr}

As mentioned above, the high ionization rate is in tension with the GeV \gray emission measured in the CMZ, which indicates another LECR component to provide the high ionization rate. In this work, we consider two scenarios for possible additional LECR nuclei accounting for the high ionization rate in the CMZ. One is the higher low-energy flux of the CRs predicted by the ``concave'' spectra in the previous theoretical nonlinear shock acceleration calculations \cite[e.g.,][]{Amato2005, Caprioli2011}. Although such effects are insignificant in other nonlinear shock acceleration calculations \cite{Caprioli2012}, such concave spectral shape can also be formed in the situation when particles are reaccelerated by multiple shocks \cite[e.g.,][]{Vieu2022}. The following calculations refer to such concave spectral shapes as scenario A.  
The other is possible anomalous CRs (ACRs), also dubbed the ``carrot'' component, accelerated from the F-, G-, and K-type stars or stellar winds from OB stars that reside in the CMZ (hereafter referred to as scenario B).  

For scenario A, we assume the continuous injection of strong-shock accelerated or reaccelerated CRs has a piecewise power-law distribution to characterize the possible concavity of the injected spectra, as represented by $Q_{\rm in}$ [Eq.\ref{equ:qin}]. Here, we set $E_{\rm b}=0.2$\,GeV, $\alpha_1=2.0$, and assume $\alpha_2$ equal to $\alpha_1$ or larger than $\alpha_1$, i.e., $\alpha_2=$2.0, 2.5, 3.0, 3.5, or 4.0, to account for the possible concave structure.  The CR proton density in the CMZ $N_{\rm lb}$ (in unit of ${\rm GeV}^{-1} {\rm cm}^{-3}$) is calculated using a simplified solution [Eq.\ref{equ:lb}] of the leaky-box model, assuming the escape time of the protons, $\tau_{\rm esc}$, is $1.5\times10^{7}{(E/1{\rm GeV})}^{-0.5}{\rm yr}\,$. Here, the energy loss time of the proton $\tau_{\rm loss}(E,n)$ is approximated by $E/P(E,n)$, in which $P(E,n)$ represents the proton energy loss rate via ionization \citep{Padovani2009} and proton-proton inelastic collision \citep{Kafexhiu2014} when propagating in gas medium with the average hydrogen density of the medium $n_{\rm H}=1\,{\rm cm}^{-3}$. The fluxes of CR proton with $E<1\,{\rm MeV}$ are set to be zero and are not taken into account for the following calculations.

For scenario B, we assume the continuous injection of strong-shock accelerated CRs has a simple power-law distribution, i.e., set $\alpha_1=\alpha_2=2.0$ for $Q_{\rm in}$, and we calculate corresponding proton density $N_{\rm lb}$ applying Eq.\ref{equ:lb}.
Then we add a local carrot component of density $N_{\rm c}(E)$ represented by Eq.\ref{equ:car}, in which $\alpha=4.0$ or 5.0. Here, we vary factor $f_{\rm c}$ for different enhancements of low-energy CRs and set $E_{\rm cut}=$1, 2, 4, and 10\,MeV, respectively.

\begin{equation}
\label{equ:qin}
Q_{\rm in}(E)=\!\! =\!\! \left\{ \begin{array}{ll} Q_{\rm 0}\left[\frac{p(E)}{p(E_{\rm b})}\right]^{-\alpha_1},\; & {\rm if}\, E \geq E_{\rm b}  \\
\\
Q_{\rm 0}\left[\frac{p(E)}{p(E_{\rm b})}\right]^{-\alpha_2},
\; & {\rm if}\, E < E_{\rm b}  \\
\end{array}\right.\!\!\!.
\end{equation}

\begin{equation}
\label{equ:lb}
N_{\rm lb}(E)=\!\! =\!\!   Q_{\rm in}(E)\left[\frac{1}{\tau_{\rm esc}(E)}+\frac{1}{\tau_{\rm loss}(E,n)}\right]^{-1}
\\
\end{equation}

\begin{equation} 
\label{equ:car}
N_{\rm c}(E)=\!\! =\!\!  \left\{ \begin{array}{ll} f_{\rm c} N_{\rm lb}(E_{\rm b})\left[\frac{p(E)}{p(E_{\rm b})}\right]^{-\alpha}, & {\rm if}\, E \geq E_{\rm cut} \\
\\
0, & {\rm if}\, E < E_{\rm cut} 
\end{array}\right.\!\!\!.
\end{equation}

\begin{figure}[h]
\centering
\includegraphics[width=0.95\columnwidth]{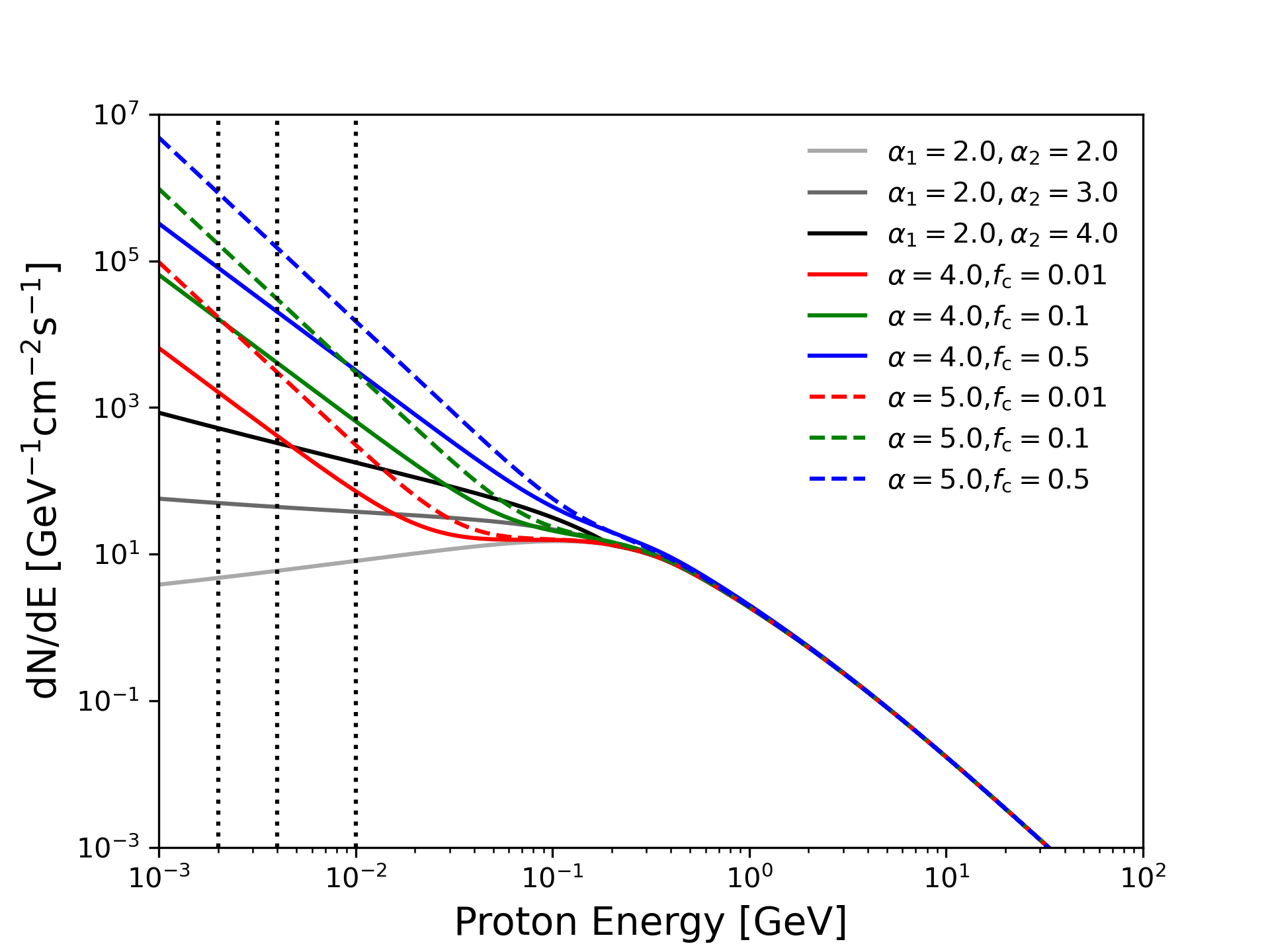}
\caption{Examples of the calculated differential flux of CR protons of scenario A with different $\alpha_1$ and $\alpha_2$ and scenario B with different $\alpha$ and $f_{\rm c}$ while setting $E_{\rm cut}=1$\,MeV. The vertical dotted lines indicate where other cutoff energies, i.e., $E_{\rm cut}=$2, 4, and 10\,MeV, are set.}   
\label{fig:cr}
\end{figure}

For HECRs in the CMZ, according to recent GeV-TeV observation results \citep{Huang2021cmz}, we fixed the proton energy density $w_{\rm p}  (80\,{\rm GeV}\leq E_{\rm p} \leq 5\,{\rm TeV}$) to be 0.045\,${\rm eV/cm}^{3}$. Assuming a distance of 8\,kpc for GC,  the calculated differential fluxes of CR protons for scenarios A and B are exemplified in Fig.\ref{fig:cr}.

HECRs and LECRs are very likely associated with different sources; thus, the elemental composition of the CRs should vary with energy. In this work, for simplicity, we apply consistent compositions for both HECRs and LECRs.
To estimate the indirect observable resulting from the interaction of CRs and the gases in the CMZ, we tested two different compositions for the CRs in the CMZ. One is the empirical GCR Source composition adopted from Table~1 of \citealt{Meyer1998} (hereafter referred to as GCRS composition), and the other is Voyager measurement of local LECR abundance listed in Table~3 of \citealt{Cummings2016} (hereafter referred to as the local composition). As for the gas content in the CMZ, we only apply solar abundances and solar abundances with twice the solar metallicity (hereafter referred to as double-solar or 2-solar composition), which is the same treatment done by \citet{Benhabiles2013}. The corresponding number density ratios relative to H of different compositions are listed in Table\ref{tab:ab}.

\begin{table}
\caption{The elemental compositions (number density ratio relative to H) applied for the calculations}
\begin{tabular}{cccc}
\hline
 & GCRS $^{\rm a}$ & Local $^{\rm b}$ & Solar $^{\rm c}$\\ 
 \hline
H& 1 & 1 & 1  \\
He&$6.90\times10^{-2}$  &$8.14\times10^{-2}$&$8.41\times10^{-2}$  \\ 
C &$3.00\times10^{-3}$ &$1.67\times10^{-3}$&$2.46\times10^{-4}$\\ 
N &$1.37\times10^{-4}$  &$2.44\times10^{-4}$&$7.24\times10^{-5}$\\ 
O &$3.72\times10^{-3}$  &$1.57\times10^{-3}$&$5.37\times10^{-4}$  \\ 
Ne &$2.82\times10^{-4}$ & $1.51\times10^{-4}$&$1.12\times10^{-4}$\\
Mg&$7.34\times10^{-4}$  &$2.26\times10^{-4}$& $3.47\times10^{-5}$ \\
Si&$7.07\times10^{-4}$  &$1.90\times10^{-4}$& $3.39\times10^{-5}$  \\
S& $9.24\times10^{-5}$ &$2.09\times10^{-5}$& $1.45\times10^{-5}$  \\
Ar&$1.52\times10^{-5}$  &$4.55\times10^{-6}$& $3.16\times10^{-6}$  \\
Ca&$4.20\times10^{-5}$  &$1.20\times10^{-5}$&  $2.04\times10^{-6}$  \\
Fe&$7.13\times10^{-4}$  &$1.15\times10^{-4}$&$2.88\times10^{-5}$\\
\hline
\end{tabular}
\\
\footnotesize{
$^{\rm a}$ The empirical GCR source composition \cite[see][Table~1]{Meyer1998}.\\
$^{\rm b}$ Voyager's measurement of local LECR abundance \cite[see][Table~3]{Cummings2016}.\\
$^{\rm c}$ The recommended present-day solar abundance \cite[see][Table~6]{Lodders2010}.\\}
\label{tab:ab}
\end{table}

\begin{figure}
\centering
\includegraphics[width=0.95\columnwidth]{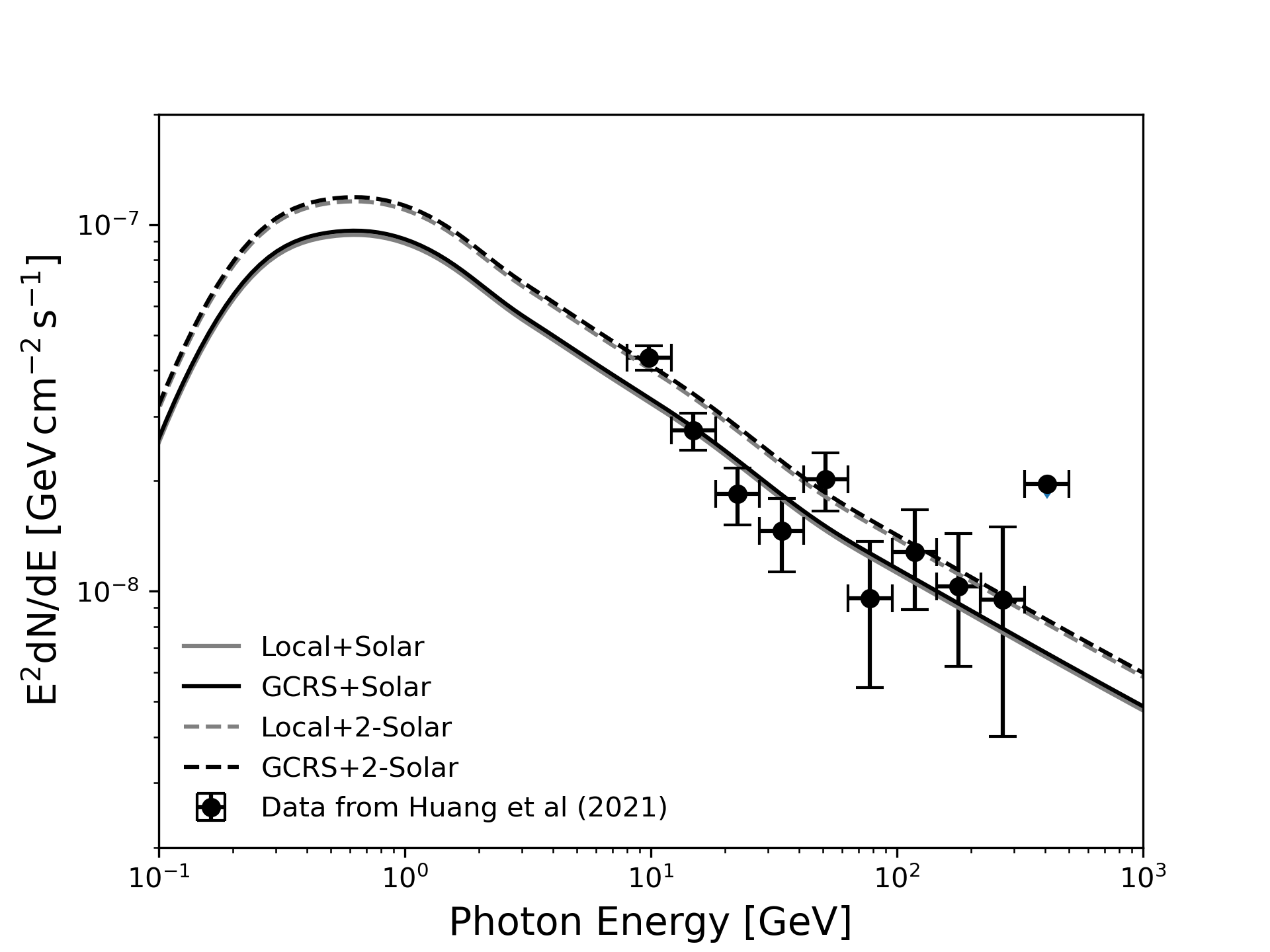}
\caption{The pion-decay emission calculated with different compositions for the CRs and gases in the CMZ when assuming a simple power-law injection ($\alpha_1=\alpha_2=2.0$) of CR protons. The observation data is from \citet{Huang2021cmz}.}
\label{fig:pion}
\end{figure}

Using a fixed total hydrogen mass, i.e., $M_{\rm H}=5\times10^{7}\,{\rm M}_{\odot}$, we then estimated the pion-decay gamma-ray emission fluxes in the CMZ  applying the parametrized cross sections from \citet{Kafexhiu2014}. As shown in Fig.\ref{fig:cr}, the fluxes of protons with energy above the energy threshold ($\sim 280\,{\rm MeV}$) of pion production process via proton-proton inelastic collision are basically the same for the tested parameters for both scenario A and scenario B. The resulting gamma-ray spectra show no obvious difference when varying the injected proton spectra parameters. When assuming different compositions of CRs and the gases, the enhancement factor calculated by Eq.(24) from \citet{Kafexhiu2014} changes, and the gamma-ray fluxes change accordingly. 
As shown in Fig.\ref{fig:pion}, the fluxes assuming a double-solar composition for gases in the CMZ are $\sim 20\%$ higher than that of solar abundance assumption; meanwhile, changing the CR composition from GCRs to local abundance measured by Voyager has minimal impact on the GeV gamma-ray fluxes. However, the predicted gamma-ray fluxes from the pion-decay process are consistent with the recent analysis results from \citet{Huang2021cmz} (data points in Fig.\ref{fig:pion}). In short, the tested $\alpha_2$ for scenario A and $f_{\rm c}$ for scenario B with the above elemental composition assumptions are consistent with current GeV-TeV gamma-ray observations in the CMZ.

\section{III. Constraint of LECRs from multi-wavelength observations towards CMZ}
\label{sec:multi}

\begin{figure}
\includegraphics[width=0.95\columnwidth]{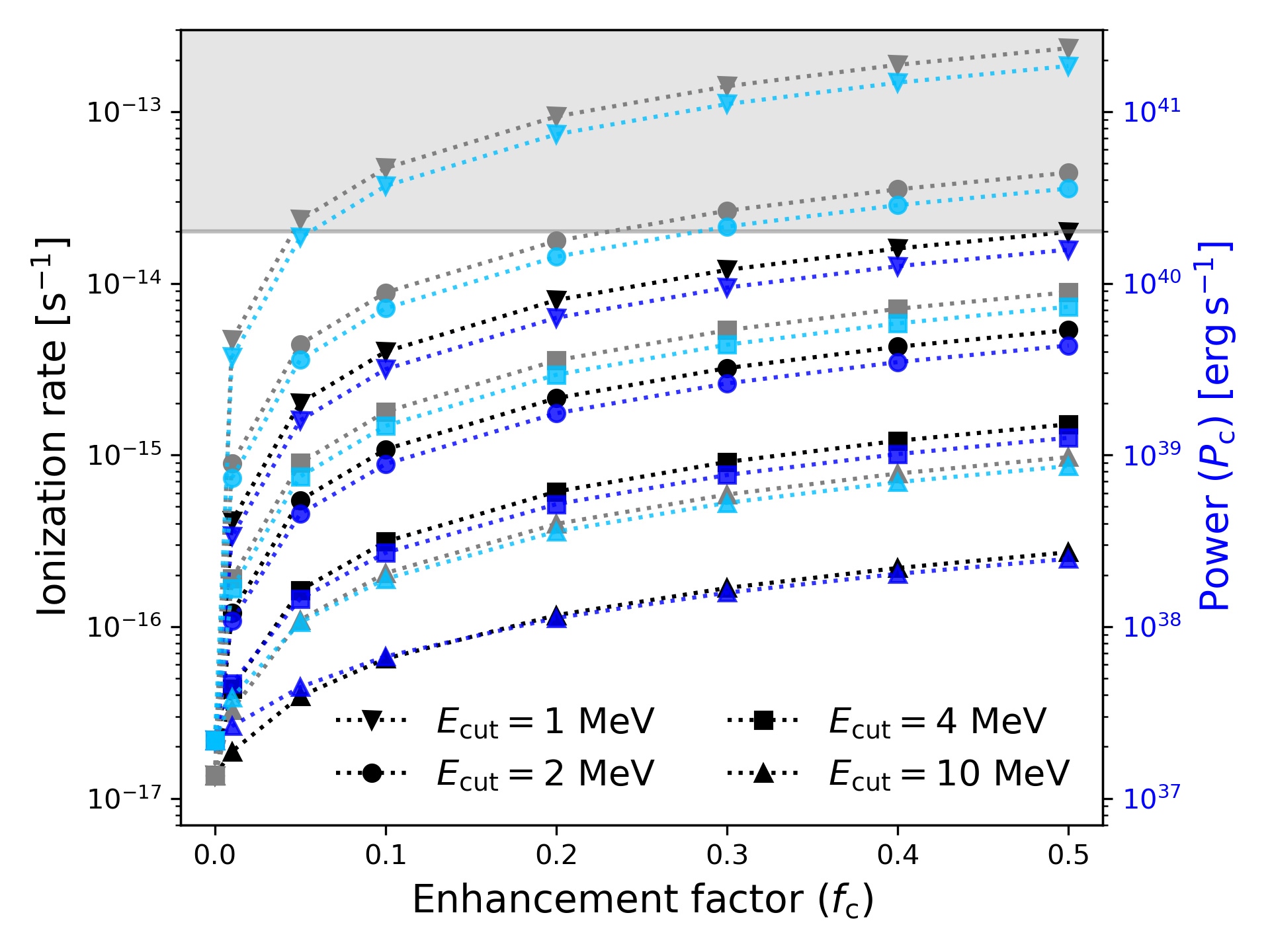}
\caption{The CR ionization rates of molecular hydrogen from the assumed local carrot component with GCRS composition and the power required to sustain the carrot component with different enhancement factors ($f_{\rm c}$). The black ($\alpha$=4.0) data and the gray ($\alpha$=5.0) data are the ionization rates, meanwhile,
the blue ($\alpha$=4.0) and light blue ($\alpha$=5.0) data represent the sustain power.}
\label{fig:irate_power}
\end{figure}

To constrain the possible LECR component in the CMZ, we then calculated the  CR ionization rate of molecular hydrogen by applying formulas in \citet{Indriolo2009} for the assumed CR spectra in Sec.2.  Different assumptions of the elemental composition will influence the calculation results, for the ionization rate of molecular hydrogen,  the results are $\sim1.5$ times higher when using GCRS composition for CRs than the results obtained by applying local measurements from the Voyagers.  For simplicity, here we only discuss the results obtained when assuming GCRS composition for CRs.
As shown in Fig.\ref{fig:irate_power}, without the additional LECRs, i.e., assuming a simple power-law injection ($\alpha_1=\alpha_2=2.0$), the CR ionization for  $E_p>1$\,MeV in the CMZ is about $1.3\times10^{-17}\, {\rm s}^{-1}$, which is three orders lower than the recent measurement ($\sim 2\times10^{-14}\, {\rm s}^{-1}$) made by \citet{Oka2019}. 
For scenario A, 
the highest ionization rate is obtained while assuming $a_2=4.0$; however, the value $\sim1.6\times10^{-16}\, {\rm s}^{-1}$ is still two orders lower than recent observation results. 
For scenario B,  as shown in Fig.\ref{fig:irate_power}, the value of $E_{\rm cut}$ has a significant influence on the calculated ionization rate, and the rate is two orders of magnitude lower when $E_{\rm cut}=10$\,MeV comparing to that of $E_{\rm cut}=1$\,MeV. Assuming $\alpha=4.0$, the ionization rate reaches $\sim2\times10^{-14}\, {\rm s}^{-1}$ when setting $f_{\rm c}=0.5$ and $E_{\rm cut}=1$\,MeV, and the corresponding total energy density of the CR protons $W_{\rm p} (E_{\rm p}>1\,{\rm MeV})$ is $0.88\, {\rm eV\,cm^{-3}}$.  Meanwhile, if $\alpha=5.0$, the lower energy cutoff ($E_{\rm cut}$) of the CR spectra should be higher than 2\,MeV for $f_{\rm c}=$0.2 and higher than $\sim 4$ MeV for $f_{\rm c}=$0.5, of which the $W_{\rm p} (E_{\rm p}>1\,{\rm MeV})<1.3\, {\rm eV\,cm^{-3}}$.   

Furthermore, we estimated the total power to sustain such an additional carrot component in the CMZ ($P_{\rm c}$) by applying Eq.\ref{equ:power} adapted from \citet{Recchia2019}, 
\begin{equation}
\label{equ:power}
P_{\rm c}= \!\! =\!\! \int E  N_{\rm c}(E) V_{\rm cmz}/\tau_{\rm loss}(E,n_{\rm cmz})dE \\
\\
\end{equation}

Here, $\tau_{\rm loss}(E,n_{\rm cmz})\propto1/n_{\rm H}\propto V_{\rm cmz}/M_{\rm H}$. The required power for different setting of $E_{\rm cut}$
and $f_{\rm c}$ is shown in Fig.\ref{fig:irate_power}. Under the constraints of the observed ionization rate, the total power to sustain such a carrot component in the CMZ is lower than $2\times10^{40}\, {\rm erg\, s}^{-1}$. This is a huge energy budget, but as mentioned above, such a low-energy ACR component is believed to be accelerated even in the low-mass F-type stars, considering the energy injection of all the stars in the CMZ, we cannot formally rule out such possibilities. 

In addition, LECR nuclei are one of the causes of Fe K$\alpha$ line emissions in the GC. Thus, to estimate the contribution from the assumed carrot component, we calculated the corresponding 6.4-keV Fe K$\alpha$ line emission using cross sections given by \citet{Tatischeff2012}. 
The total line fluxes are all lower than $10^{-4}$\,${\rm ph\,cm}^{-2}\,{\rm s}^{-1}$ for the tested CR spectra even assuming double-solar abundance for the gases in the CMZ. The highest line flux is about $2\times10^{-5}$\,${\rm ph\,cm}^{-2}\,{\rm s}^{-1}$, if the ionization rate caused by the additional LECR component $\leq 2\times10^{-14}\, {\rm s}^{-1}$.
The 6.4-keV line flux caused by the possible additional LECR nuclei in the CMZ
is below the observed total flux in the CMZ region \cite{Uchiyama2013}. 
Since the Fe K$\alpha$ line emission has multiple origins,  such as hard X-ray photoionization and collisional ionization induced by CR electrons, our results are consistent with current X-ray observations toward the GC.

\section{IV. MeV nuclear deexcitation line emission in the CMZ}
\label{sec:nlines}

\begin{figure}
\includegraphics[width=0.95\columnwidth]{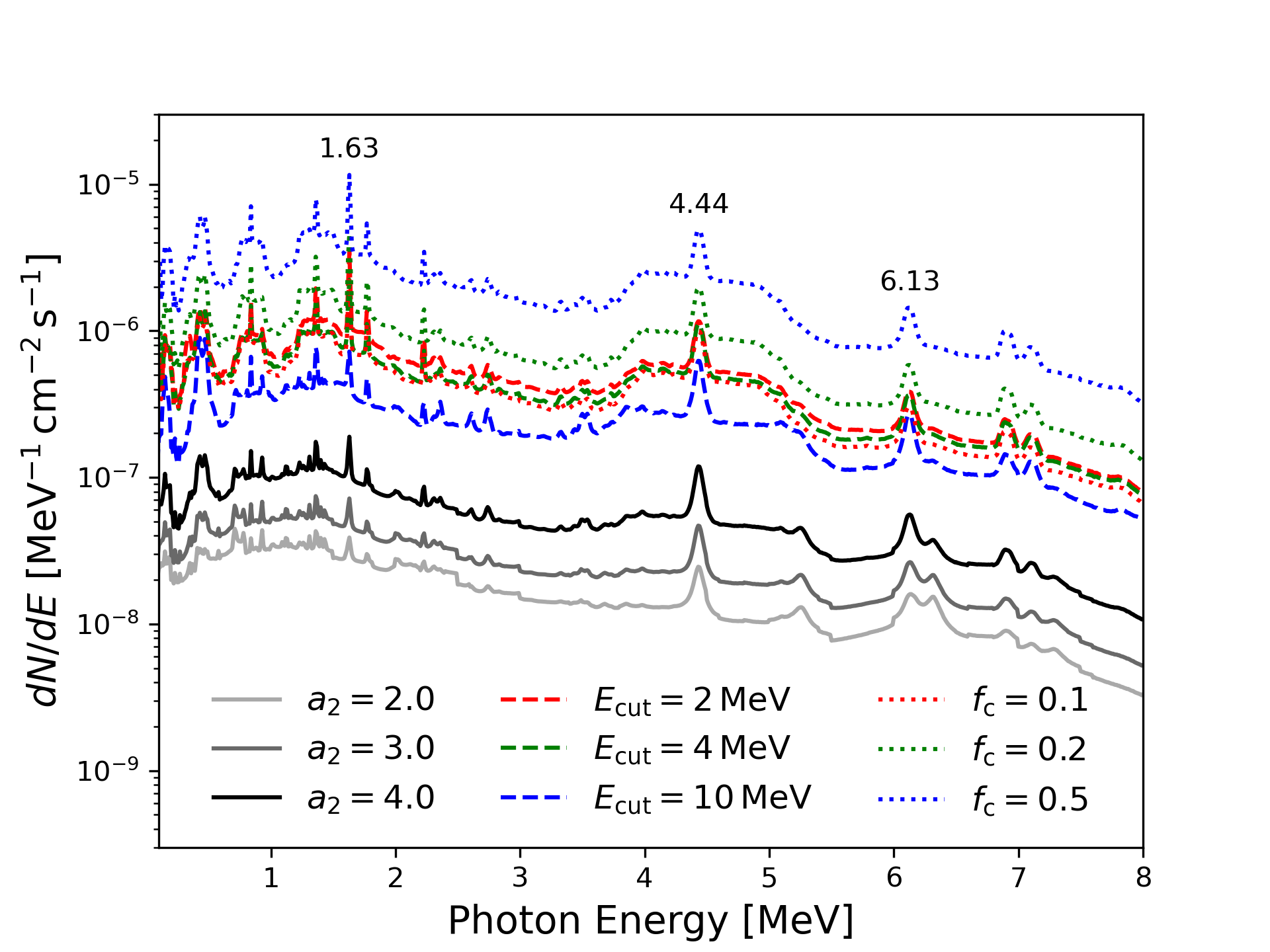}
\caption{Examples of the calculated MeV nuclear deexcitation line emission of scenario A and scenario B.
Solid lines: scenario A with different $\alpha_2$ while setting $\alpha_1=2.0$.  
Dashed lines: scenario B with different $E_{\rm cut}$ while setting $\alpha=4.0$ and $f_{\rm c}=0.5$. Dotted lines: scenario B with different $f_{\rm c}$ while setting $\alpha=5.0$ and $E_{\rm cut}=4.0$\,MeV.  The compositions of CR nuclei and gases in the CMZ are assumed to be GCRS and 2-solar respectively. 
}
\label{fig:lines}
\end{figure}

\begin{figure}
\includegraphics[width=0.9\columnwidth]{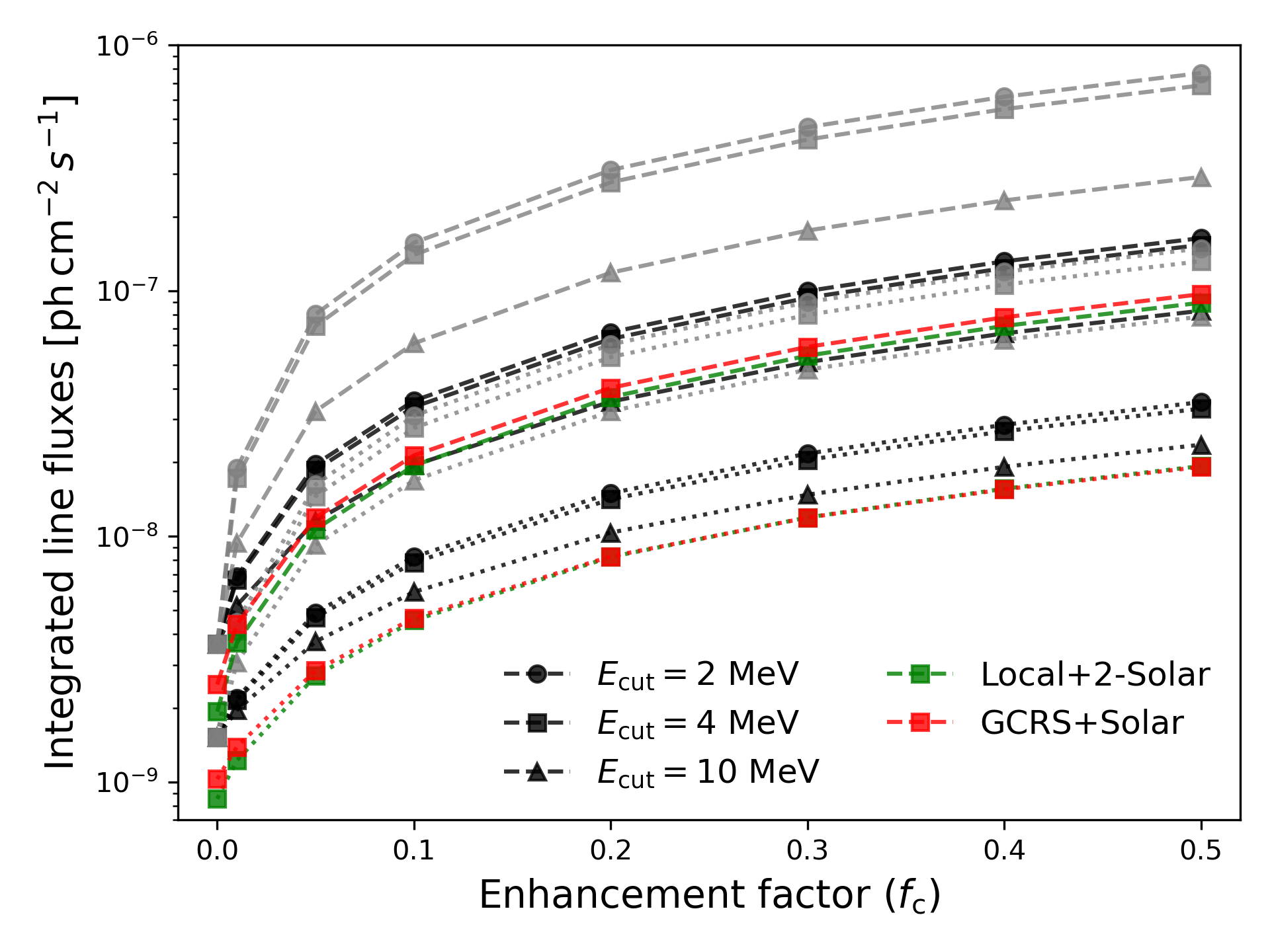}
\caption{The integrated fluxes of 4.44-MeV line (dashed lines) and 6.13-MeV line (dotted lines) from the CMZ under the assumption of different fractions of the carrot component $f_{\rm c}$ and $E_{\rm cut}$. The black data show the results assuming $\alpha=4.0$ and the grey data are obtained assuming $\alpha=5.0$, assuming the compositions of CR nuclei and gases in the CMZ are assumed to be GCRS and 2-solar respectively. Setting $E_{\rm cut}=4$\,MeV, the green data are obtained from the local (for the CRs)+2-solar (for the gases) assumption; meanwhile, the  red data are obtained from the GCRS (for the CRs)+solar (for the gases) assumption.}
\label{fig:lineflux}
\end{figure}

Under the constraint of ionization rate measurements, we calculated the possible nuclear deexcitation line emission from the interaction between CR nuclei and the gases towards the CMZ. Both narrow \gray lines from excited heavier elements of the ambient gas that collide with CR protons and $\alpha$ particles and broadened \gray lines from excited heavy nuclei of LECRs that interact with hydrogen and helium of the ambient gas are taken into account during the calculation.  For the calculation process, except for the code TALYS which is the newest 1.96 version \cite{talys2008,Koning2014}, we applied the same procedure as described in Section 3.1 of \citet{nlines}, which followed the method developed by \citet{Ramaty1979,Murphy2009,Benhabiles2013}. 
Assuming GCRS and double-solar compositions for CR nuclei and gases in the CMZ respectively, the calculated differential nuclear deexcitation line emission flux for both scenario A and scenario B assuming different spectral parameters are exemplified in Fig.\ref{fig:lines}. As shown in Fig.\ref{fig:lines}, strong line emission can be found at energies such as 1.63, 4.44, and 6.13 MeV, which is mainly produced via the deexcitation of excited $^{20}$Ne, $^{12}$C, and $^{16}$O. 
Using the above calculated MeV emission line spectra, we then estimated the 4.44- and 6.13-MeV line fluxes integrated for a linewidth of $\sim100$\,keV. For scenario A, the highest fluxes, $\sim 1.44 \times10^{-8} {\rm \,cm^{-2}\,s^{-1}}$ for 4.44-MeV line and $\sim 4.53 \times10^{-9} {\rm \,cm^{-2}\,s^{-1}}$ for 6.13-MeV line, are obtained when assuming $\alpha_2=4.0$.
Such fluxes are 5 times higher than a simple power-law assumption for the continuously injected CRs in the CMZ ($\alpha_1=\alpha_2=2.0$).
For scenario B, the line fluxes assuming different enhancements of additional carrot components and low-energy cutoffs are illustrated in Fig.\ref{fig:lineflux}.  Within the constraint of ionization rate ($\zeta \leq 2\times10^{-14} {\rm s}^{-1}$), assuming $\alpha=4.0$, the highest fluxes of the 4.44-MeV line and the 6.13-MeV line, $\sim 1.6 \times10^{-7} {\rm \,cm^{-2}\,s^{-1}}$ and $\sim 3.5\times10^{-8} {\rm \,cm^{-2}\,s^{-1}}$ respectively, are obtained when setting $E_{\rm cut}\leq 4\,{\rm MeV}$ and $f_{\rm c}=0.5$. Meanwhile, assuming $\alpha=5.0$, the 4.44-MeV line and the 6.13 MeV line fluxes can reach $\sim 6.9 \times10^{-7} {\rm \,cm^{-2}\,s^{-1}}$ and $\sim 1.3\times10^{-7} {\rm \,cm^{-2}\,s^{-1}}$ respectively when setting $E_{\rm cut}=4\,{\rm MeV}$ and $f_{\rm c}=0.5$. Moreover, assuming a local composition for the CRs or solar abundance for the gases will yield lower fluxes for these line emissions, as also exemplified by the red and yellow data in Fig.\ref{fig:lineflux}.

\begin{figure}
\centering
\includegraphics[width=0.9\columnwidth]{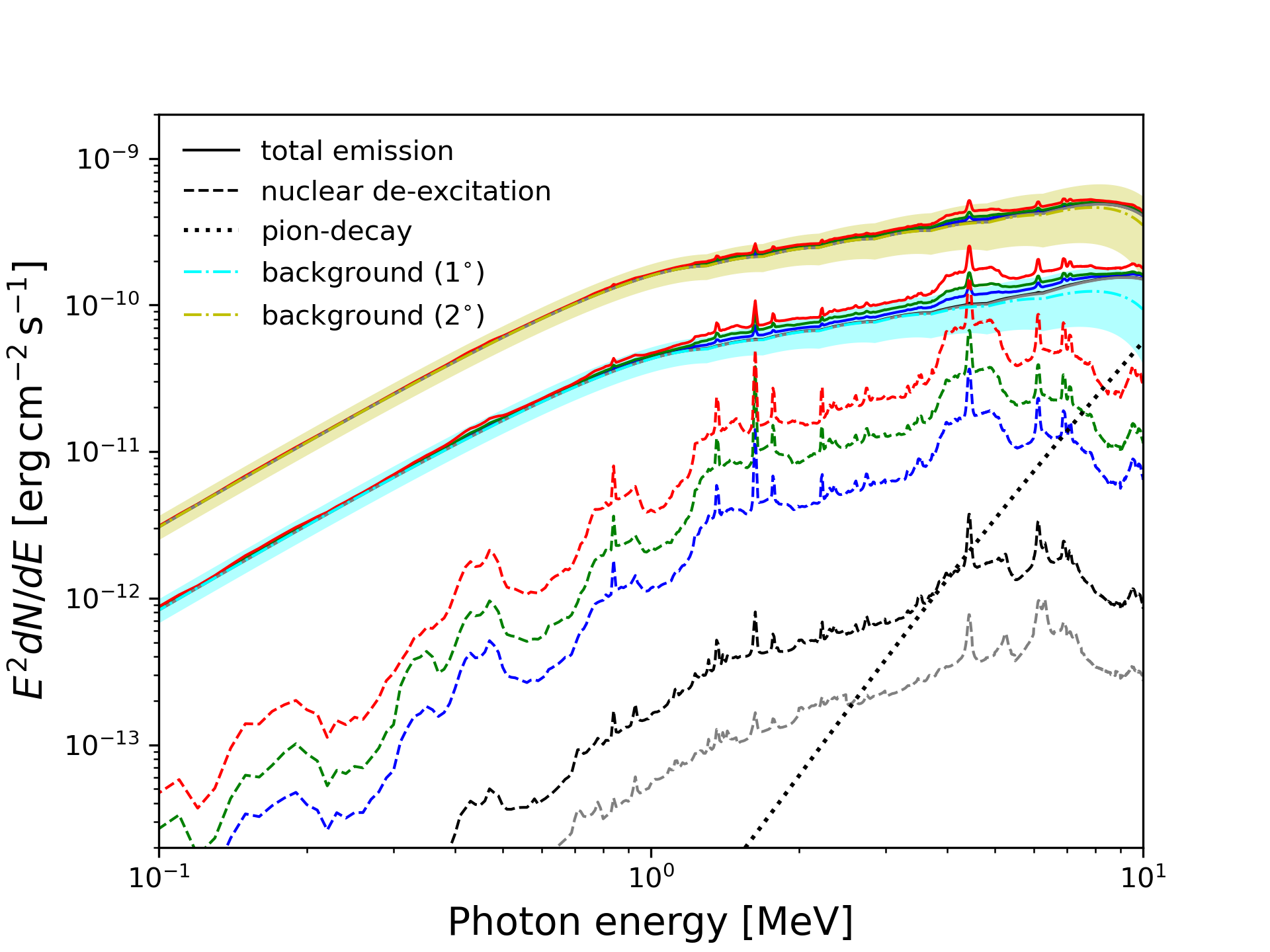}
\caption{Possible MeV gamma-ray emission from the CMZ with different scenarios of CR spectral distribution. The total emission (solid lines) includes emission from nuclear deexcitation (dashed lines) and pion-decay (dotted line) processes, as well as the continuum MeV background within different radii around the GC (dashed-dotted lines) derived by INTEGRAL/SPI data \cite{Siegert2022}. The shaded areas represent possible diffuse background emission with $1\sigma$ uncertainty. Red lines: $\alpha=5.0$, $f_{\rm c}=0.5$, and $E_{\rm cut}=4\,{\rm MeV}$ for scenario B. Green lines: $\alpha=5.0$, $f_{\rm c}=0.2$, and $E_{\rm cut}=2\,{\rm MeV}$ for scenario B. Blue lines: $\alpha=4.0$, $f_{\rm c}=0.5$, and $E_{\rm cut}=1\,{\rm MeV}$ for scenario B.  Black lines: $\alpha_1=2.0$ and $\alpha_2=4.0$ for scenario A. Gray lines: $\alpha_1=\alpha_2=2.0$ for scenario A. The compositions of CR nuclei and gases in the CMZ are assumed to be GCRS and 2-solar respectively.
}
\label{fig:bkg}
\end{figure}

To discuss the possible detection of MeV line emission towards the CMZ, we must consider the influence of the MeV continuum background, which cannot be ignored in the GC region. 
Recently, using 16 years of INTEGRAL/SPI observation in the band 0.5-8.0 MeV,  \citet{Siegert2022} fitted the Galactic diffuse emission in a region of $\Delta l \times \Delta b = 95^{\circ}\times 95^{\circ}$ around the GC with energy-dependent IC scattering emission template acquired from the GALPROP code \citep{galprop}, and found the spectrum of the diffuse gamma rays can be well described by a power law with an index of $\sim-1.39 $.  Based on the energy-dependent spatial template and the spectrum of Galactic diffuse emission obtained by \citet{Siegert2022},  we extracted the diffuse gamma-ray emission spectrum within a radius of $2^{\circ}$ around the GC, as the possible continuum MeV background when next-generation MeV telescopes with an angular resolution of $\sim2^{\circ}$ observing the CMZ. As shown in Fig.\ref{fig:bkg}, the derived MeV diffuse gamma-ray flux within the $2^{\circ}$ region around the GC is much higher than the MeV deexcitation line emission obtained from the assumed spectra of LECRs in the CMZ. Such a high background flux will make the MeV line research towards the CMZ very difficult.  However, if the angular resolution of the next-generation MeV
gamma-ray detectors could be improved to better than $2^{\circ}$ at $\lesssim 10$\,MeV, $1^{\circ}$ for example, the chance of future detection of the deexcitation line emission from the CMZ will be significantly improved.  
As exemplified in Fig.\ref{fig:bkg}, the derived flux of the background MeV continuum for a region within a radius of $1^{\circ}$ from GC is at the same level of the MeV deexcitation line emission obtained for extreme assumptions of LECRs in the CMZ, i.e., the high ionization rate is fully caused by the CR nuclei.  
Moreover, future detectors will resolve more MeV gamma-ray sources, thus, the background flux will be lower than our current estimation. Thus, we have a great opportunity to constrain these theories of the LECR distribution in the CMZ through the observation of MeV deexcitation line emission via next-generation MeV gamma-ray detectors.

\section{V. Summary}
\label{sec:sum}

The extremely high ionization rate as high as $\sim2\times10^{-14}\, {\rm s}^{-1}$ is observed in CMZ. The most probable ionizer in the dense region in CMZ is the LECRs. However, the GeV-TeV observations in CMZ reveal a similar  HECR density in the CMZ as in the solar neighborhood, which implies an ionization rate 3 orders of magnitude smaller than observed. Thus, an additional LECR component is required. In this work, we consider two possible origins of the additional LECR component: one is from the concavity of the CR accelerated by strong shocks due to nonlinear effects or re-acceleration effects, and the other is the very soft carrot component accelerated from the F-, G-, and K-type stars or stellar winds from OB stars that reside in the CMZ. We found that only the latter can explain the extremely high ionization rate observed. If the CR ionization rate $\sim2\times10^{-14}\, {\rm s}^{-1}$, we found that for $\alpha=4.0$ and $E_{\rm cut}=1$\,MeV, an enhancement factor [$f_{\rm c}$ in Eq.\ref{equ:car}] of 0.5 is required  and if $\alpha=5.0$, then $E_{\rm cut}\geq 2$\,MeV for $f_{\rm c}=0.2$ and $E_{\rm cut}\geq 4$\,MeV for $f_{\rm c}=0.5$.  
A total LECR energy budget of about $2\times10^{40}\, {\rm erg\, s}^{-1}$ is required in the above extreme case, which is huge but not impossible.  

Suppose additional LECR nuclei components mainly cause the high ionization rate in the CMZ. In that case, the predicted fluxes of the strong nuclear deexcitation lines can reach a level of $\sim 1\times10^{-6} {\rm ph\, cm}^{-2}\,{\rm s}^{-1}$ for 4.44-MeVline and  $>1\times10^{-7} {\rm ph\, cm}^{-2}\,{\rm s}^{-1}$ for 6.13-MeV line, which may be detected by the next-generation MeV instruments such as COSI \cite{cosi}, AMEGO \cite{amego} and MeGaT. However, the chemical composition of both LECRs and gases can considerably impact the MeV results. Due to the high continuum background in the GC region, both good spectral and angular resolution are required to detect these MeV deexcitation lines. In particular, the width of the nuclear deexcitation lines can be as high as several percent ($\sim 100~\rm keV$) mainly due to the kinematics \cite{Ramaty1979}.  Thus the angular resolutions play a more important role in detecting such kinds of MeV emissions.  In this regard, the next-generation MeV instruments with an angular resolution similar to or better than $1^{\circ}$ are very promising to test such a hypothesis and solve the mysterious origin of the high ionization rate in the CMZ. Such in situ measurement of the LECR spectrum would also help to understand the possible dynamic effects of the LECR component in the CMZ as well as the possible impact on the star-formation process in this complex region. 

\section{Data Availability}

To calculate emissivities of the deexcitation \gray line lines,  we used the code TALYS  (version 1.96,\citealt{talys2008}), which could be downloaded from {\url{https://tendl.web.psi.ch/tendl_2019/talys.html}. For a better match with the experiment data, we modified the deformation files of $^{14}$N, $^{20}$Ne, and $^{28}$Si using the results of \citet{Benhabiles2011}. 
We also used the production cross sections of the specific lines listed in the compilation of \citet{Murphy2009}.

\section{Acknowledgements}

Bing Liu acknowledges the support from the NSFC under grant 12103049. Rui-zhi Yang is supported by the NSFC under grants 12041305, and 12393854. 



\appendix
\bibliography{cmz.bib} 

\label{lastpage}

\end{document}